# New mechanism for EM waves generation as a response to earthquakes

A. Haddadou and R. Annou

Department of theoretical physics
Faculty of physics –USTHB (Algiers)-ALGERIA

## Abstract

A mechanism that explains the increase of electromagnetic (EM) perturbations during earthquakes is proposed. When earthquakes occur, surface waves propagate along the globe surface inducing a globe surface rippling, which is reproduced in an amplified manner at the atmosphere-ionosphere interface. A surface EM wave that imparts a quiver velocity to electrons propagates along the rippled interface encountering a periodic plasma density. A non-linear density current is consequently generated that is a source of EM waves. The wave equation is solved and the EM field is determined.

When a stored energy of a chemical, gravitational or simply an elastic form is released suddenly in the earth, an earthquake occurs. It is manifested at our level by the shaking of the soil. The shaking is a response to elastic waves that travel through the rocks. However, for the facture of the rocks to occur, the accumulated stress ought to exceed the strength of these rocks. Earthquakes may be induced as well, by some human activities, such as, detonation of large underground explosions, and filling of large reservoirs. Around the fracture, seismic waves are generated and radiated in all directions. These waves may be classified into volume and surface waves. The volume waves that propagate within the earth contain the primary waves (p), which are longitudinal and capable to propagate in solids as well as in liquids, and the secondary waves (s), which are transverse and are transmitted only in solids by way of shearing of rocks[1-3]. On the other hand, the surface waves that are confined to propagate along the earth surface, consist of Love and Rayleigh waves. Both the waves induce horizontal motion, nonetheless only Rayleigh waves are a source of vertical soil movement. In fact, at distances greater that 100 km form the source, the main damage is primarily due to surface waves. During the deformation, rocks exhibit an increase in volume, which affects consequently the seismic velocities, electric resistivity, as well as soil and water levels. Furthermore, the coupling between the ground in motion and the atmosphere leads to an increase of wave processes activity. For instance, hours before an earthquake, a green light attributed to atomic oxygen is radiated[1,4]. It is confirmed experimentally also, that in connection with earthquakes, electromagnetic waves are generated in the ionosphere[5]. Besides, by way of ionosondes and GPS receivers, it has been measured an increase of the variability of electron concentration in the ionosphere[6]. The mechanism at the root of the transformation of acoustic waves generated at the globe surface, in EM waves in the ionosphere has not been satisfactorily elucidated. To this end, a satellite called Demeter (Detection of EM Emissions Transmitted from Earthquake Regions) has been recently launched by CNES (the French Centre National des études Spatiales)[7]. Nonetheless, there was an attempt to shine some light on this issue in Ref.[8], where a collisional model is suggested. We believe however, that this model does not reproduce an actual picture of the wave transformation mechanism, since it overlooks the effect of the atmosphere-ionosphere interface physics. We propose in this note a model that takes into account the coupling between the soil movement and the atmosphere, as well as the physics of the interface.

Prior to any analysis, let's expose the main idea of our theory related to the generation of EM waves as a response to earthquakes. Any rippling of the soil due to the propagation of a seismic surface wave such as a Rayleigh wave is transmitted to the ionosphere by way of an acoustic wave that propagates in the atmosphere. As a consequence, the interface between the neutral atmosphere and the ionized ionosphere gets rippled. It may be recalled, that periodical structures are induced by powerful laser irradiation on metallic surfaces that enhance the energy transfer through coupling with surface electromagnetic waves[9-11]. Consequently we dwell an analogy with the problem of laser/metal interaction cited above, and model the rippled atmosphere/ionosphere interface by a periodic electron density (c.f. Ref.[12]). The equation of the interface may be written as follows,

$$x_i = a \cos q (z - \frac{\Omega}{q}) \tag{1}$$

However, at a sharp discontinuity in plasma density a surface wave exists and propagates along the $\hat{z}$-axis. Its electric field $\vec{E} = \vec{A}(x)e^{-i(\omega t - kz)}$, satisfies the wave equation,

$$\frac{d^2 E}{dt^2} - \alpha^2 E = 0, \tag{2}$$

where, $\alpha^2 = \alpha_I^2 = k^2 - \frac{\omega^2}{c^2}$ for x<0 and $\alpha^2 = \alpha_{II}^2 = k^2 - \frac{\omega^2}{c^2}\varepsilon$ for x>o, where $\varepsilon = 1 - \frac{\omega_p^2}{\omega^2}$ for x>0 (=1 for x<0) is the permittivity of the medium, and $\omega_p$ is the plasma frequency. A regular solution of Eq. (2) at infinity that is also continuous at x=0, is given by,

$$A(x) = A e^{\alpha_I x} \quad (x<0), \tag{3-a}$$

$$= A e^{-\alpha_{II} x} \quad (x>0), \tag{3-b}$$

The dispersion relation of the surface wave may be obtained by using Maxwell's equations along with the continuity of the normal component of the displacement vector at x=0. It is cast as,

$$k^2 = \frac{\omega^2}{c^2} \frac{\varepsilon}{1+\varepsilon} \tag{4}$$

In our case, the surface wave does not propagate along the $\hat{z}$ axis in a homogeneous plasma, but rather in an inhomogeneous plasma with a periodic density. During its propagation, the surface wave encounters matter on a segment of $\frac{p}{q}$ length, and vacuum on the subsequent one, in an alternating manner. Hence, the plasma density may be modeled by the following function,

$$\frac{n(z)}{n_0} = \begin{cases} 0, & \text{for } \frac{p}{q} < z - \frac{\Omega}{q}t < 0 \\ \\ 1, & \text{for } 0 < z - \frac{\Omega}{q}t < \frac{p}{q} \end{cases} \quad (5)$$

The density may be expanded in a Fourier series,

$$\frac{n(z)}{n_0} = \sum C_m e^{i m q [z - \frac{\Omega}{q}t]}, \quad (6)$$

where, $C_0 = 1/2$ and $C_m = \frac{1}{2 p i m}(1 - e^{-i m p})$. The component of the first order is being given by $\frac{n_1}{n_0} = \frac{1}{ip} e^{iq(z - \frac{\Omega}{q}t)}$.

Moreover, the surface wave imparts oscillatory velocity to free electrons,

$$\vec{v} = \frac{e \vec{E}}{m i w}, \quad (7)$$

and produces a non-linear current density $\vec{J}_{NL}$ at $(w+\Omega, \vec{q}+\vec{k})$,

$$\vec{J}_{NL} = -e n_1 \vec{v}$$

$$= \frac{e^2 n_0}{p m w} \exp[-i\{(w+\Omega)t - (k+q)z\}]\vec{A}(x) \quad (8)$$

The presence of a current in the plasma is a source of electromagnetic waves. Indeed, the wave equation governing the propagation of electromagnetic waves, if one takes into account the current density is given by[13,14],

$$\nabla^2 \vec{e} - \nabla(\nabla \cdot \vec{e}) - \frac{1}{c^2}\frac{\partial^2 \vec{D}}{\partial t^2} = \frac{4p}{c^2}\frac{\partial \vec{J}_{NL}}{\partial t} \quad (9)$$

The y- component of the electromagnetic $\vec{e}$ is then governed by the reduced relation,

$$\nabla^2 e_y + \frac{w^2}{c^2} e e_y = P e^{-a_s |x|} \times e^{-i(w+\Omega)t} \times e^{i(k+q)z} \qquad (10)$$

where, $P = -\frac{i w_p^2}{c^2}(1+\frac{\Omega}{w})\frac{A_y}{p}$ and $s = I, II$.

Let us consider changing the variables, i.e., $e_y = e'_y e^{i(k+q)z}$, then one gets,

$$\frac{\partial^2 e'_y}{\partial x^2} + \Gamma^2 e'_y = P e^{-a_s |x|} e^{-i(w+\Omega)t} \qquad (11)$$

where, $\Gamma^2 = \frac{w^2}{c^2} e - (k+q)^2$.

The solution of Eq. (11) is given by,

$$e'_y(x) = \begin{cases} C_2 e^{-i\Gamma_0 x} + \dfrac{P e^{a_I x} e^{-i(w+\Omega)t}}{a_I^2 + \Gamma_0^2} & (x<0) \\ \\ C_1 e^{+i\Gamma x} + \dfrac{P e^{-a_{II} x} e^{-i(w+\Omega)t}}{a_{II}^2 + \Gamma^2} & (x>0) \end{cases} \qquad (12)$$

To determine the constants $C_1$ and $C_2$, we impose the continuity of the solution (12) at x=0, and integrate Eq(11) across x=0 as well. The constants are then given by,

$$C_1 = \frac{P e^{-i(w+\Omega)t}}{i(\Gamma+\Gamma_0)}\left\{\frac{a_{II} - i\Gamma_0}{a_{II}^2 + \Gamma^2} + \frac{a_I + i\Gamma_0}{a_I^2 + \Gamma_0^2}\right\},$$

and (13)

$$C_2 = \frac{P e^{-i(w+\Omega)t}}{i\Gamma_0}\left\{\frac{\Gamma_0^2 + i\Gamma_0 a_{II}+1}{a_{II}^2 + \Gamma^2} + \frac{-\Gamma_0^2 + i\Gamma_0 a_I + 1}{a_I^2 + \Gamma_0^2}\right\}.$$

To conclude we recall, that before and during earthquakes electromagnetic waves are reported. These disturbances that occur before the shaking, have been studied earlier, and thought to be a tool for earthquakes prediction. We present in this note a new mechanism to explain the EM waves' generation during the earthquakes. As such, the phenomenon is not relevant for prediction, but we believe it could be a diagnostic tool. We state that when an earthquake occurs, surface waves propagate along the globe surface, and the variation of the soil level induces an acoustic wave that propagates in the atmosphere up to the ionosphere. Hence, the ripple of the soil is reproduced at the level of the atmosphere/ ionosphere interface. After having developed the model, we fortunately, found an earlier work that gives justifications to our hypothesis. Indeed, it is stated in Ref.[15] that Rayleigh waves induce atmospheric pressure disturbances by vertical pulsed action. The waves may even propagate up to heights of $\sim$ 110 Km without noticeable damping for the low frequency components. Moreover, the amplitude increases with increasing altitude. The soil ripple amplitude is magnified at the atmosphere-ionosphere interface. Furthermore, it is established that at the interface of vacuum/plasma, a surface wave propagates. Now that the ripple is formed, we show that the SEW (surface EM wave) encounters a periodic plasma density. The SEW imparts to electrons. As a consequence a non-linear current density vector is generated, which is a source of EM waves.